\documentclass[letterpaper,12pt]{article}
\usepackage{amssymb}
\usepackage{amsmath}
\usepackage{mathptmx}
\usepackage{scalefnt}

\input{tcilatex}
\begin{document}

\noindent George Karabatsos* and Stephen G. Walker\medskip \smallskip 
\newline
\begingroup%
\scalefont{1.4}%
\textbf{A} \textbf{Bayesian Nonparametric Hypothesis Testing Approach for
Regression Discontinuity Designs}\footnote{%
This research is supported in part by NSF grant SES-1156372.}%
\endgroup%

\begin{center}
February 7, 2014

\bigskip
\end{center}

\noindent \textbf{\noindent Abstract}: \ The regression discontinuity (RD)\
design is a popular approach to causal inference in non-randomized studies.
This is because it can be used to identify and estimate causal effects under
mild conditions. Specifically, for each subject, the RD design assigns a
treatment or non-treatment, depending on whether or not an observed value of
an assignment variable exceeds a fixed and known cutoff value.

In this paper, we propose a Bayesian nonparametric regression modeling
approach to RD designs, which exploits a local randomization feature. In
this approach, the assignment variable is treated as a covariate, and a
scalar-valued confounding variable is treated as a dependent variable (which
may be a multivariate confounder score). Then, over the model's posterior
distribution of locally-randomized subjects that cluster around the cutoff
of the assignment variable, inference for causal effects are made within
this random cluster, via two-group statistical comparisons of treatment
outcomes and non-treatment outcomes.

We illustrate the Bayesian nonparametric approach through the analysis of a
real educational data set, to investigate the causal link between basic
skills and teaching ability.

\textbf{\ }\ \textbf{\bigskip }

\newline
\textbf{Short title}:\ \ Bayesian Nonparametric Hypothesis Testing for\
RD.\newpage

\section{\textbf{Introduction}}

A basic objective in scientific research is to infer causal effects of a
treatment versus non-treatment from empirical data. Randomized studies
provide a gold standard of causal inference [19], because such a study
ensures, through the random assignments of treatments to subjects, that
treatment assignments are independent of all observed and unobserved
confounding variables which characterize experimental subjects. As a
consequence, the causal effect can be estimated by a direct comparison of
treatment outcomes and non-treatment outcomes. However, in many research
settings, it is not feasible to randomly assign treatments to subjects; due
to financial, ethical, or time constraints [19]. Therefore, in such
settings, it becomes necessary to estimate causal effects of treatments from
non-randomized, observational designs.

For observational studies, a popular approach to causal inference is based
on the regression discontinuity (RD)\ design ([21], [7]). In the design,
each subject is assigned to a treatment or non-treatment whenever an
observed value of a continuous-valued assignment variable ($R$) exceeds a
cutoff value. The RD design can provide a "locally-randomized experiment",
in the sense that the causal effect of treatment outcomes versus
non-treatment outcomes can be identified and estimated from subjects having
values of the assignment variable that are located in a small neighborhood
around the cutoff. As shown in [11], the RD\ design can empirically produce
causal effect estimates that are similar to those from a standard randomized
study ([1], [5], [3], [2], [20]). Arguably, the fact that RD designs can
provide such locally-randomized experiments is a main reason why, between
1997 through 2013, that at least 74 RD-based empirical studies have emerged
in the fields of education, psychology, statistics, economics, political
science, criminology, and health sciences ([7], [16], [4], [24], [17]).

Given the locally-randomization feature of RD designs, a
conceptually-attractive approach to causal inference is to identify the
cluster 
\begin{equation*}
\mathcal{C}_{\epsilon }(r_{0})=\{i:|r_{i}-r_{0}|<\epsilon \}
\end{equation*}%
of locally-randomized subjects, who have assignment variable observations $%
r_{i}$ that are located in a neighborhood of size $\epsilon >0$ around the
cutoff $r_{0}$. Then, for subjects within the cluster $\mathcal{C}_{\epsilon
}(r_{0})$, to perform statistical tests by comparing treatment outcomes
between subjects with $r_{i}\geq r_{0}$ against non-treatment subjects; i.e.
among subjects with $r_{i}<r_{0}$) [6].

One approach to identifying the locally-randomized cluster of subjects is to
iteratively search for the largest value of $\epsilon >0$ that leads to a
non-rejection of the null hypothesis tests of zero effect of the treatment
variable $T=\mathbf{1}(R\geq r_{0})$. A theoretical motivation for this
approach is that, as a consequence of local randomization, the distribution
of the confounding variables is the same for non-treatment subjects located
just to the left of the cutoff ($r_{0}$), as for subjects located just to
the right of the cutoff [15].

However, this approach is not fully satisfactory.\ This is because it bases
estimates and tests of causal effects (comparisons of treatment outcomes
versus non-treatment outcomes), conditionally on an optimal value of the
neighborhood size parameter $\epsilon $, i.e., the clustering configuration $%
\mathcal{C}_{\epsilon }(r_{0})$ found via the null hypothesis tests
mentioned above. Therefore, the approach does not fully account for the
uncertainty that is inherent in the neighborhood size parameter, i.e., the
configuration of the cluster of locally-randomized subjects. As a result,
the approach may lead to null-hypothesis tests of causal effects that have
exaggerated significance levels.

In this manuscript, we propose a Bayesian nonparametric regression approach
to estimating and testing for causal effects, for locally-randomized
subjects in a RD design, while accounting for the uncertainty in the cluster
configuration $\mathcal{C}_{\epsilon }(r_{0})$ of locally-randomized
subjects. In the approach, we consider a scalar variable $X$ that we assume
sufficiently describes all observed and unobserved pre-treatment
"confounding" variables. Again, as a consequence of local-randomization, $X$
has the same distribution, for subjects located just to the left and for
subjects located just to the right of the cutoff, respectively [15].

Then, we fit a Bayesian nonparametric regression model, based on the
restricted Dirichlet process (rDP)\ [23], in order to provide a flexible
regression of $X$ on the assignment variable $R$. Importantly, the model
provides a random clustering of subjects, with respect to common values of $%
X $, in a way that is sensitive to the ordering of the values of $R$. Given
a posterior random draw of clusters of subjects under the rDP\ model, we
identify the single cluster 
\begin{equation*}
\mathcal{C}(r_{0})=\{i\,:\,\mbox{individual}\,\,i\,\,\mbox{has an}%
\,\,r_{i}\,\,\mbox{in the cluster containing}\,\,r_{0}\}
\end{equation*}%
of locally-randomized subjects. Within this cluster we then compare the
treatment outcomes (of subjects with $r_{i}\geq r_{0}$) against
non-treatment outcomes (of subjects with $r_{i}<r_{0}$), via statistical
summaries and two-sample statistical tests. Specifically, we may compare
outcomes in terms of the mean, variance, quantiles, and the interquartile
range, and use various two-sample tests, including the $t$-test,
Wilcoxon-Mann-Whitney test of equality of medians, chi-square test of
equality of variances, and the Kolmogorov-Smirnov test. We can even estimate
the probability $\Pr [Y_{1}\geq Y_{0}|\mathcal{C}(r_{0})]$ that the
treatment outcome ($Y_{1}$) exceeds the non-treatment outcome ($Y_{0}$) (see
[14]). We then average the results of such statistical comparisons over many
posterior samples of the clusters $\mathcal{C}(r_{0}),$under the restricted
DP model. Hence, when making such statistical comparisons, we fully account
for the uncertainty that is inherent in the clustering configuration $%
\mathcal{C}(r_{0})$ of locally-randomized subjects. Moreover, this
statistical procedure represents another application of inference of
posterior functionals in DP-based models [8].

We can extend the Bayesian nonparametric regression procedure to RD design
setting that involve multiple pre-treatment confounding variables $%
(X_{1},\ldots ,X_{p})$. In this case, we can construct a scalar confounding
covariate $X$ by the multivariate confounder score [18], which enables the
comparison of treatment outcomes and non-treatment outcomes, conditionally
on subclassified values of the score. Multivariate confounder scores are
constructed by a regression of the outcomes $Y$ on $(T,X_{1},\ldots ,X_{p})$%
, and then setting each score to be equal to the part of the predictor that
is free of the linear effect of $T$ on $Y$. Moreover, while in this paper we
base our causal inference procedure on the restricted DP, in fact, it is
possible to base the causal inference procedure on any Bayesian
nonparametric regression model that can cluster subjects as a function of
the assignment variable $R$. See [13], for a recent example.

In the next section, we review the Bayesian nonparametric regression model,
based on the rDP. Then we further describe how estimates and tests of the
causal effects is undertaken based on the clustering method. We also provide
some more details on the multivariate confounder scoring method. Moreover,
we show how this Bayesian nonparametric method can be extended to handle
causal inferences in a context of a fuzzy RD\ design [22], which involves
imperfect treatment compliance among the subjects. In Section 3, we
illustrate our model through the analysis of a data set, to provide causal
inferences in an educational research setting. Section 4 ends with
conclusions.

\section{The Bayesian Nonparametric \textbf{Model}}

Let $\{(x_{i},r_{i},y_{i})\}_{i=1}^{n}$ denote a sample set of data obtained
under a RD\ design. For each subject $i\in \{1,\ldots ,n\}$, the triple $%
(x_{i},r_{i},y_{i})$ denotes an observed value of the confounding variable,
the assignment variable, and outcome variable, respectively. For such data,
we consider a Bayesian nonparametric regression of the pre-treatment
confounding variable $X$ on the assignment variable $R$, based on a
restricted DP\ (rDP) mixture of normal linear regressions [23]. Without loss
of generality, we assume 
\begin{equation*}
r_{1}\leq r_{2}\leq \cdots \leq r_{n}.
\end{equation*}%
The key idea is that clusterings will be based on this order; so the first
cluster would be based on the smallest values of $r$, and so on. Since here
we are interested in clustering subjects on common values of $X$, based on
values of $R$, we represent this model as a random partition model, as
follows: 
\begin{subequations}
\label{restDP}
\begin{align}
\left[ (X_{1},\ldots ,X_{n})|r,\rho _{n},\{(\beta _{j}^{\ast },\sigma
_{j}^{2\ast })\}_{j=1}^{k_{n}}\right] \text{ }& \sim
\dprod\limits_{j=1}^{k_{n}}\dprod\limits_{\{i:s_{i}=j\}}^{{}}\mathrm{Normal}%
\bigg(x_{i}|\underline{\mathbf{r}}_{i}^{\intercal }\boldsymbol{\beta }%
_{j}^{\ast },\sigma _{j}^{2\ast }\bigg) \\
\lbrack \rho _{n}]\text{ }& \sim \text{ }\pi (\rho _{n})=\dfrac{\alpha
^{k_{n}}}{\alpha ^{\lbrack n]}}\dfrac{n}{k_{n}!}\dprod\limits_{j=1}^{k_{n}}%
\dfrac{1}{n_{j}}\mathbf{1}\left\{ s_{1}\leq \cdots \leq s_{n}\right\} \\
\lbrack \boldsymbol{\beta }_{j}|\sigma _{j}^{2}]& \sim \mathrm{Normal}\big(%
\boldsymbol{\beta }_{j}|\beta _{0},\sigma _{j}^{2}\mathbf{C}^{-1}\big) \\
\lbrack \sigma _{j}^{2}]& \sim \mathrm{InverseGamma}\big(\sigma _{j}^{2}|a,b%
\big),
\end{align}%
where $\underline{\mathbf{r}}_{i}=(1,r_{i})^{\intercal }$; the collection $%
\{(\boldsymbol{\beta }_{j}^{\ast },\sigma _{j}^{2\ast })\}_{j=1}^{k_{n}}$
form the $k_{n}\leq n$ distinct values in the sample of parameters $((%
\boldsymbol{\beta }_{1},\sigma _{1}^{2}),\ldots ,(\boldsymbol{\beta }%
_{n},\sigma _{n}^{2}))$ that are assigned to each of the $n$ subjects. And $%
\rho _{n}=(s_{1},\ldots ,s_{n})$ is a random partition of the $n$
observations, where $s_{i}=j$ if $(\boldsymbol{\beta }_{i},\sigma _{i}^{2})=(%
\boldsymbol{\beta }_{j}^{\ast },\sigma _{j}^{2\ast })$, and with \noindent $%
n_{j}=\tsum\nolimits_{i=1}^{n}\mathbf{1}\{(\boldsymbol{\beta }_{i},\sigma
_{i}^{2})=(\boldsymbol{\beta }_{j}^{\ast },\sigma _{j}^{2\ast })\}$.

Also, the rDP is parameterized by a precision parameter $\alpha $, and by a
normal-inverse-gamma baseline distribution for $(\boldsymbol{\beta },\sigma
^{2})$ that defines the mean of the process. See [23] for more details.

The posterior distribution of the clustering partition $\rho
_{n}=(s_{1},\ldots ,s_{n})$, with respect to the $k_{n}$ distinct values $\{(%
\boldsymbol{\beta }_{j}^{\ast },\sigma _{j}^{2\ast })\}_{j=1}^{k_{n}}$, is
given by: 
\end{subequations}
\begin{equation*}
\pi (\rho _{n}|\mathbf{y},\mathbf{x})\propto \frac{\alpha k_{n}}{k_{n}!}%
\dprod\limits_{j=1}^{k_{n}}\frac{1}{n_{j}}\sqrt{\dfrac{|\mathbf{C}|}{|%
\mathbf{C}+\underline{\mathbf{R}}_{j}^{\intercal }\underline{\mathbf{R}}_{j}|%
}}\dfrac{b^{a}\Gamma (a+n_{j}/2)}{\Gamma (a)(b+V_{j}^{2}/2)^{a+n_{j}/2}}%
\mathbf{1}_{s_{1}\leq \cdots \leq s_{n}}.
\end{equation*}%
In the above, 
\begin{equation*}
V_{j}^{2}=(\mathbf{y}_{j}-\widehat{\mathbf{y}}_{j})\widehat{\mathbf{W}}_{j}(%
\mathbf{y}_{j}-\widehat{\mathbf{y}}_{j}),
\end{equation*}%
\begin{equation*}
\widehat{\mathbf{W}}_{j}=\mathbf{I}_{j}-\mathbf{R}_{j}^{\intercal }(\mathbf{C%
}+\underline{\mathbf{R}}_{j}^{\intercal }\underline{\mathbf{R}}_{j})_{j}^{-1}%
\underline{\mathbf{R}}_{j}^{\intercal }
\end{equation*}%
and 
\begin{equation*}
\widehat{\mathbf{y}}_{j}=\underline{\mathbf{R}}_{j}\boldsymbol{\beta }_{0},
\end{equation*}%
with $\underline{\mathbf{R}}_{j}$ the matrix of row vectors $\underline{%
\mathbf{r}}_{i}=(1,r_{i})^{\intercal }$ for subjects belonging in cluster
group $j$ [23]. Posterior samples from $\pi (\rho _{n}|\mathbf{y},\mathbf{x}%
) $ can be generated though the use of a reversible-jump Markov Chain Monte
Carlo (RJMCMC) sampling algorithm that is described in Section 4 of [23]. At
each sampling stage of the algorithm, with equal probability, either two
randomly-selected clusters of subjects, that are adjacent with respect to
the ordering of $R$, are merged into a single cluster; or a randomly
selected cluster of subjects is split into two clusters.

Given a random sample of a partition, $\rho _{n}\sim \pi (\rho _{n}|\mathbf{y%
},\mathbf{x})$ from the posterior, and given the fixed index $i\equiv i_{0}$
of a subject whose assignment variable $r_{i}$ is nearest to the cutoff $%
r_{0}$, we then identify the single (posterior random)\ cluster of
locally-randomized subjects, with this cluster of subjects identified by the
subset of indices 
\begin{equation*}
\mathcal{C}(r_{0})=\{i:s_{i_{0}}=s_{i}\}.
\end{equation*}
This is a posterior random cluster of subjects with values of the assignment
variables $r_{i}$ located in a neighborhood around the cutoff $r_{0}$.

For the subjects in this random cluster, we compare treatment outcomes $%
y_{i} $ for subjects where $r_{i}\geq r_{0}$, versus non-treatment outcomes $%
y_{i}$ for subjects where $r_{i}<r_{0}$, based on two-sample statistical
comparisons of various statistical quantities (e.g., means), as mentioned in
Section 1. We then repeat this process over a large number of MCMC\ samples
from the posterior distribution of the partitions, $\pi (\rho _{n}|\mathbf{y}%
,\mathbf{x})$.\ We then summarize the posterior distribution of such
statistical comparisons over these samples, in order to provide estimates
and tests of causal effect of the treatment versus non-treatment, on the
outcome $Y$.

This procedure can be extended to a fuzzy RD design, where the assignment
variable $R$ represents eligibility to receive a treatment, and some
subjects who are assigned treatment $T_{i}=\mathbf{1}(r_{i}\geq r_{0})$ opt
to receive the other treatment $(1-T_{i})$. Then, for the subset of subjects
in a given random cluster $\mathcal{C}_{\epsilon }(r_{0})$, we can divide
the difference in statistical quantities (e.g., treatment mean minus
non-treatment mean) by the difference $\overline{T}_{\mathcal{C}%
}^{(r_{i}\geq r_{0})}-\overline{T}_{\mathcal{C}}^{(r_{i}<r_{0})}$, where $%
\overline{T}_{\mathcal{C}}^{(r_{i}\geq r_{0})}$ ($\overline{T}_{\mathcal{C}%
}^{(r_{i}<r_{0})}$, respectively) is the average $T_{i}$ for subjects in the
locally-randomized cluster $\mathcal{C}(r_{0})$ that have assignment
variables $r_{i}\geq r_{0}$ (with assignment variables $r_{i}<r_{0}$,
respectively). Such a divided difference provides an instrumental-variables
estimate of a causal effect of treatment versus non-treatment.\ This is
true, provided that the local exclusion restriction holds in the sense that
for a given cluster of subjects $\mathcal{C}_{\epsilon }(r_{0})$, any effect
of the assignment variable $\mathbf{1}(R\geq r_{0})$ on $Y$\ must be only
via $T$ [16].

For an RD setting that involves multiple pre-treatment confounding variables 
$\mathbf{x}_{i}=(x_{1i},\ldots ,x_{pi})^{\intercal }$ (for $i=1,\ldots ,n$),
we construct scalar-valued confounding variable $x_{i}$ ($i=1,\ldots ,n$) by
taking Miettinen's multivariate confounder score [18], with $x_{i}=\widehat{%
\beta }_{0}+\widehat{\boldsymbol{\beta }}_{\mathbf{x}}g(\mathbf{x}_{i})$,
based on 
\begin{equation*}
\widehat{\mathrm{E}}[Y_{i}|\mathbf{x}_{i},r_{i}]=\widehat{\beta }_{0}+%
\widehat{\boldsymbol{\beta }}_{\mathbf{x}}B(\mathbf{x}_{i})^{\intercal }+%
\widehat{\beta }_{R}\mathbf{1}(r_{i}\geq r_{0}),
\end{equation*}%
with $B(\mathbf{\cdot })$ a chosen (e.g., polynomial)\ basis transformation
of $\mathbf{x}$, and with coefficient estimates 
\begin{equation*}
\widehat{\boldsymbol{\beta }}=(\widehat{\beta }_{0},\widehat{\boldsymbol{%
\beta }}_{\mathbf{x}},\widehat{\beta }_{R})
\end{equation*}%
obtained by a linear model fit. We consider the Bayesian estimator 
\begin{equation*}
\widehat{\boldsymbol{\beta }}=(v^{-1}\mathbf{I}_{q}+\mathbf{B}^{\intercal }%
\mathbf{B})^{-1}\mathbf{B}^{\intercal }\mathbf{y},
\end{equation*}%
with $\mathbf{y}=(y_{1},\ldots ,y_{n})^{\intercal }$, and with $\mathbf{B}$
the $(n\times q)$-dimensional basis matrix with row vectors $(1,B(\mathbf{x}%
_{i})^{\intercal },\mathbf{1}(r_{i}\geq r_{0}))$, $i=1,\ldots ,n$.

\begin{table}[H] \centering%
\begin{tabular}{lcc}
\hline
\textbf{Statistic} & \textbf{Non-Treatment} & \textbf{Treatment} \\ \hline
sample size & 103.1 \ (3, 190) & 6.7\ \ (2, 16) \\ 
mean & .37 \ ($-$.07, 1.55) & 1.23 \ (.97, 1.59) \\ 
variance & .76 \ (.01, 1.04) & .47 \ (.01, 0.85) \\ 
interquartile range & 1.17\ \ (.18, 1.71) & .90 \ (.24, 1.41) \\ 
skewness & $-$.11\ \ (-1.26, .71) & .03 \ ($-$.63, 0.82) \\ 
kurtosis & 2.69\ \ (1.45, 3.41) & 2.06 \ (1.00, 3.06) \\ 
1\%ile & $-$1.34\ \ ($-$2.20, 1.47) & .27 \ ($-$0.65, 1.47) \\ 
10\%ile & $-$.77 \ ($-$1.35, 1.47) & .42 \ (.01, 1.47) \\ 
25\%ile & $-$.20 \ ($-$0.65, 1.47) & .74 \ (.30, 1.47) \\ 
50\%ile & .34 \ ($-$.18, 1.47) & 1.22 \ (1.00, 1.59) \\ 
75\%ile & .98 \ (.53, 1.65) & 1.65 \ (1.47, 2.06) \\ 
90\%ile & 1.43 \ (1.24, 1.71) & 2.14 \ (1.71, 2.77) \\ 
99\%ile & 2.04 \ (1.71, 2.42) & 2.28 \ (1.71, 2.89) \\ 
t-statistic & \multicolumn{2}{c}{$-$2.02 \ ($-$4.21, .88) \ \ p-value:\ \
.19 \ (.00, .91)} \\ 
$F$ test,variance & \multicolumn{2}{c}{\ 4.86 \ (.02, 34.45) \ \ p-value:\
.65\ \ (.05, .98)} \\ 
$\Pr [Y_{1}\geq Y_{0}|\mathcal{C}_{\epsilon }(r_{0})]$ & \multicolumn{2}{c}{
0.70 (.21, .93)} \\ 
$\Pr [Y_{1}\leq Y_{0}|\mathcal{C}_{\epsilon }(r_{0})]$ & \multicolumn{2}{c}{
0.22 (.04, .67)} \\ 
KS test & \multicolumn{2}{c}{.28\ (.05, .98)} \\ \hline
\end{tabular}%
\caption{Statistical comaprisons of treatment outcomes versus non-treatment
outcomes.}\label{TableKey}%
\end{table}%

\section{\textbf{Illustration}}

A data set was obtained under a partnership between four Chicago University
schools of education, which implemented a new curriculum that aims to train
and graduate teachers to improve Chicago public school education. This data
set involves $n=204$ undergraduate teacher education students, each of whom
enrolled into one of the four Chicago schools of education during either the
year of 2010, 2011, or 2012 (90\%\ female; mean age = $22.5$, s.d. = $5.3$, $%
n=203$); $47\%$, $21\%$, $10\%$, and $22\%$ attended the four universities; $%
49\%$, $41\%$ and $10\%$\ enrolled in 2010, 2011, and 2012). We investigate
the causal effect of basic skills on teacher performance (e.g., [10]),
because most U.S. schools of education based their undergraduate admissions
decisions on the ability of individual applicants to pass basic skills
tests. Here, the assignment variable is a 4-variate random variable, defined
by subtest scores on an Illinois test of basic skills, in reading, language,
math and writing. Each subtest has a minimum passing score of 240. The
dependent variable is the total score on the 50-item Haberman Teacher
Pre-screener assessment, and a score in the 40-50 range indicates a very
effective teacher. This assessment has a test-retest reliability of .93, and
has a 95\% accuracy rate in predicting which teachers will stay and succeed
in the teaching profession, and is used by many schools to assess applicants
of teaching positions [12]. Among all the $205$ students of the RD design,
the average Haberman Pre-screener score is $29.82$ (s.d. = $4.3$).\ The
average basic skills score in reading ($\mathrm{Read}$), language ($\mathrm{%
Lang}$), math ($\mathrm{Math}$), and writing ($\mathrm{Write}$) was $204.3$
(s.d. = $33.4$), $204.0$ (s.d. = $35.8$), $212.6$ (s.d. = $42.2$), and $%
238.3 $ (s.d. = $23.7$), respectively.

Using the Bayesian clustering method that was described in the previous
section, based on the rDP regression model, we analyzed the data set to
estimate the causal effect of passing the reading basic skills exam
(treatment), versus not passing (non-treatment), on students' ability to
teach in urban schools, for subjects located around the cutoff $r_{0}$\ of
an assignment variable $R$. We treated the Haberman z-score as the outcome
variable $Y$. Also, using the Bayesian rDP regression model, we regressed
the scalar-valued confounding variable $X$ on $R$. The confounding variable $%
X$ is a multivariate confounder score constructed from 113 pre-treatment
variables that describe students' personal background, high school
background, and teaching preferences. The multivariate confounder score is
based on the the Bayesian linear model fit with prior parameter $v=1000$,
and with linear basis $B(\mathbf{x})=(1,\mathbf{x})^{\intercal }$, as
described in the previous section. Also, the assignment variable $R$ is
defined by $\mathrm{B240d10}=(\min (\mathrm{Read,Lang,Math,Write})-240)/10$%
.\ This gives a minimum difference between the four basic subtest scores
subtracted by the minimum cutoff score (240), and provides one standard
method for handling a multivariate assignment variable [24]. Then, the
treatment assignment variable $T$ is defined by \textrm{BasicPass} $=\mathbf{%
1}(\mathrm{B240d10}\geq 0)$.

Using code we wrote in the MATLAB\ (Natick, VA) software language, we
analyzed the data set using the Bayesian rDP model. For the model, we chose
vague prior specifications $\beta _{0}=\mathbf{0},\mathbf{C}=\mathrm{diag}%
(10^{3},10)$, $\alpha =a=b=1$. All posterior estimates, reported here, are
based on 200,000\ MCMC\ samples of clusters $\mathcal{C}_{\epsilon }(r_{0})$%
, which led to accurate posterior estimates of two-group statistical
comparisons, according to standard MCMC\ convergence assessment criteria
[9]. Specifically, univariate trace plots displayed good mixing of model
parameters and posterior predictive samples, while all posterior predictive
estimates obtained 95\%\ MC\ confidence intervals with half-width sizes of
.00.

Table 1 presents the results of the two-group statistical comparisons, in
terms posterior mean and 95\% posterior credible interval summaries of the
group sample size, mean, variance, interquartile range, skewness, kurtosis,
quantiles, the t-statistic, the F-statistic for equality of variances,
exceedance probabilities $\Pr [Y_{1}\geq Y_{0}|\mathcal{C}_{\epsilon
}(r_{0})]$ and $\Pr [Y_{1}\leq Y_{0}|\mathcal{C}_{\epsilon }(r_{0})]$ of the
treatment outcome ($Y_{1}$) and the non-treatment outcome ($Y_{0}$), and the
Kolmogorov-Smirnov test for the equality of distributions. We find that, in
terms of the posterior mean of these statistics, the treatment group in the
random clusters $\mathcal{C}_{\epsilon }(r_{0})$ (having assignment variable
observations $\mathrm{B240d10}\geq 0$) tended to have higher values of the
Haberman z-score outcome $Y$ compared to the non-treatment group in the
random clusters $\mathcal{C}_{\epsilon }(r_{0})$ (having assignment variable
observations $\mathrm{B240d10}\geq 0$), in terms of the mean and quantiles.
In contrast, the outcomes of the non-treatment tended to have larger
dispersion (variance and interquartile range), and more skewness and
kurtosis, compared to the treatment group. The treatment group had a
significantly higher 90\%ile (.90 quantile) of the z-score outcome $Y$
compared to the non-treatment group, as the 95\% posterior credible
intervals of the outcome for the two groups was (1.71, 2.77) and (1.24,
1.71), respectively.

\section{\textbf{Discussion\label{Section: Discussion}}}

In this paper, we proposed and illustrated a novel, Bayesian nonparametric
regression modeling approach to RD\ designs, which exploits the local
randomization feature of RD designs, and which basis causal inferences on
comparisons of treatment outcomes and non-treatment outcomes within
posterior random clusters of locally-randomized subjects. The approach can
be easily extended to fuzzy RD\ settings, involving treatment
non-compliance. We illustrate the Bayesian nonparametric approach through
the analysis of a real educational data set, to investigate the causal link
between basic skills and teaching ability. Finally, the approach assumes
that the RD\ design provides data on all confounding variables (that are
used to construct $X$), but this assumption of no hidden bias is
questionable. Therefore, in future research, it would be of interest to
extend the procedure, so that it can provide an analysis of the sensitivity
of causal inference, with respect to varying degrees of hidden biases, i.e.,
of effects to hypothetical unobserved confounding variables.\bigskip 

\bigskip 

\noindent {\LARGE References\smallskip }

\begin{description}
\item \lbrack 1] Aiken, L., S. West, D. Schwalm, J. Carroll, and S. Hsiung
(1998): \textquotedblleft Comparison of a randomized and two
quasi-experimental designs in a single outcome evaluation efficacy of a
university-level remedial writing program,\textquotedblright\ \textit{%
Evaluation Review}, \textit{22}, 207--244.

\item \lbrack 2] Berk, R., G. Barnes, L. Ahlman, and E. Kurtz (2010):
\textquotedblleft When second best is good enough: A comparison between a
true experiment and a regression discontinuity
quasi-experiment,\textquotedblright\ \textit{Journal of Experimental
Criminology}, \textit{6}, 191--208.

\item \lbrack 3] Black, D., J. Galdo, and J. Smith (2005): \textquotedblleft
Evaluating the regression discontinuity design using experimental
data,\textquotedblright\ Unpublished manuscript.

\item \lbrack 4] Bloom, H. (2012): \textquotedblleft Modern regression
discontinuity analysis,\textquotedblright\ \textit{Journal of Research on
Educational Effectiveness}, \textit{5}, 43--82.

\item \lbrack 5] Buddelmeyer, H. and E. Skoufias (2004): \textit{An
evaluation of the performance of regression discontinuity design on PROGRESA}%
, World Bank Publications.

\item \lbrack 6] Cattaneo, M., B. Frandsen, and R. Titiunik (2013):
\textquotedblleft Randomization inference in the regression discontinuity
design: An application to the study of party advantages in the U.S.
Senate,\textquotedblright\ Technical report, Department of Statistics,
University of Michigan.

\item \lbrack 7] Cook, T. (2008): \textquotedblleft Waiting for life to
arrive: A history of the regression discontinuity design in psychology,
statistics and economics,\textquotedblright\ \textit{Journal of Econometrics}%
, \textit{142}, 636--654.

\item \lbrack 8] Gelfand, A. and A. Kottas (2002): \textquotedblleft A
computational approach for full nonparametric Bayesian inference under
Dirichlet process mixture models,\textquotedblright\ \textit{Journal of
Computational and Graphical Statistics}, \textit{11}, 289--305.

\item \lbrack 9] Geyer, C. (2011): \textquotedblleft Introduction to
MCMC,\textquotedblright\ in S. Brooks, A. Gelman, G. Jones, and X. Meng,
eds., \textit{Handbook of Markov Chain Monte Carlo}, Boca Raton, FL: CRC,
3--48.

\item \lbrack 10] Gitomer, D., T. Brown, and J. Bonett (2011):
\textquotedblleft Useful signal or unnecessary obstacle? The role of basic
skills tests in teacher preparation,\textquotedblright\ \textit{Journal of
Teacher Education}, \textit{62}, 431--445.

\item \lbrack 11] Goldberger, A. (2008/1972): \textquotedblleft Selection
bias in evaluating treatment effects: Some formal
illustrations,\textquotedblright\ in D. Millimet, J. Smith, and E. Vytlacil,
eds., \textit{Modelling and evaluating treatment effects in economics},
Amsterdam: JAI Press, 1--31.

\item \lbrack 12] Haberman, M. (2008): \textit{The Haberman Star Teacher
Pre-Screener,} Houston: The Haberman Educational Foundation.

\item \lbrack 13] Karabatsos, G. and S. Walker (2012b): \textquotedblleft
Adaptive-modal Bayesian nonparametric regression,\textquotedblright\ \textit{%
Electronic Journal of Statistics}, \textit{6}, 2038--2068.

\item \lbrack 14] Kotz, S., Y. Lumelskii, and M. Pensky (2003): \textit{The
Stress-Strength Model and its Generalizations}, New Jersey: World Scientific.

\item \lbrack 15] Lee, D. (2008): \textquotedblleft Randomized experiments
from non-random selection in U.S. house elections,\textquotedblright\ 
\textit{Journal of Econometrics}, \textit{142}, 675--697.

\item \lbrack 16] Lee, D. and T. Lemieux (2010): \textquotedblleft
Regression discontinuity designs in economics,\textquotedblright\ \textit{%
The Journal of Economic Literature}, \textit{48}, 281--355.

\item \lbrack 17] Li, F., A. Mattei, and F. Mealli (2013): \textquotedblleft
Bayesian inference for regression discontinuity designs with application to
the evaluation of Italian university grants,\textquotedblright\ Technical
report, Department of Statistics, Duke University.

\item \lbrack 18] Miettinen, O. (1976): \textquotedblleft Stratification by
a multivariate confounder score,\textquotedblright\ \textit{American Journal
of Epidemiology}, \textit{104}, 609--620.

\item \lbrack 19] Rubin, D. (2008): \textquotedblleft For objective causal
inference, design trumps analysis,\textquotedblright\ \textit{The Annals of
Applied Statistics}, \textit{2}, 808--840.

\item \lbrack 20] Shadish, W., R. Galindo, V. Wong, P. Steiner, and T. Cook
(2011): \textquotedblleft A randomized experiment comparing random and
cutoff-based assignment.\textquotedblright\ \textit{Psychological Methods}, 
\textit{16}, 179-191.

\item \lbrack 21] Thistlewaite, D. and D. Campbell (1960): \textquotedblleft
Regression-discontinuity analysis: An alternative to the ex-post facto
experiment,\textquotedblright\ \textit{Journal of Educational Psychology}, 
\textit{51}, 309--317.

\item \lbrack 22] Trochim,W. (1984): \textit{Research design for program
evaluation: The regression discontinuity approach}, Newbury Park, CA: Sage.

\item \lbrack 23] Wade, S., S. Walker, and S. Petrone (2013, to appear):
\textquotedblleft A predictive study of Dirichlet process mixture models for
curve fitting,\textquotedblright\ \textit{Scandinavian Journal of Statistics}%
, n/a--n/a.

\item \lbrack 24] Wong, V., P. Steiner, and T. Cook (2013):
\textquotedblleft Analyzing regression-discontinuity designs with multiple
assignment variables: A comparative study of four estimation
methods,\textquotedblright\ \textit{Journal of Educational and Behavioral
Statistics}, \textit{38}, 107--141.
\end{description}

\end{document}